\documentclass[article, aps, 12pt, floatfix, a4paper]{revtex4-2}
\usepackage{gensymb}
\usepackage{color}
\usepackage[normalem]{ulem}
\usepackage{graphicx}
\usepackage{dcolumn}
\usepackage{float}
\usepackage{bm}
\usepackage{amsmath}
\usepackage{comment}

\newcommand{\be}{\begin{equation}}
\newcommand{\ee}{\end{equation}}
\newcommand{\bfig}{\begin{figure}}
\newcommand{\efig}{\end{figure}}

\newcommand{\YIn}{Y$_8$CoIn$_3$}
\newcommand{\RX}{\textit{R}$_8$Co\textit{X}$\kern-0.15em_{3}$}
\newcommand{\RGa}{\textit{R}$_8$CoGa$_3$}
\newcommand{\RIn}{\textit{R}$_8$CoIn$_3$}
\newcommand{\NaBi}{Na$_3$Bi}

\begin{document}
\title{
Material realization of spinless, covalent-type Dirac semimetals \\in three dimensions 
}
\author{Yuki Tanaka$^{1}$}
\author{Rinsuke Yamada$^{1}$}\email{ryamada@ap.t.u-tokyo.ac.jp}
\author{Manabu Sato$^{1}$}
\author{Motoaki Hirayama$^{2}$}
\author{Max Hirschberger$^{1,2}$}
\email{hirschberger@ap.t.u-tokyo.ac.jp}
\affiliation{$^{1}$Department of Applied Physics and Quantum-Phase Electronics Center (QPEC), The University of Tokyo, Bunkyo, Tokyo 113-8656, Japan}
\affiliation{$^{2}$RIKEN Center for Emergent Matter Science (CEMS), Wako, Saitama 351-0198, Japan}

\maketitle

\begin{center}
\Large{Abstract}
\end{center}

\textbf{Realization of a three-dimensional (3D) analogue of graphene has been a central challenge in topological materials science. Graphene is stabilized by covalent bonding unlike conventional spin-orbit type 3D Dirac semimetals (DSMs). 
In this study, we demonstrate the material realization of covalent-type 3D DSMs \RX{} stabilized by covalent bonding. We observe that the carrier mobility $\bm{\mu}$ of Dirac fermions reaches 3,000$\,$cm$^2$/Vs even in polycrystalline samples, and $\bm{\mu}$ increases with the inverse of the Fermi energy, evidencing significant contributions to charge transport from Dirac electrons. \RX{} provides a material platform for exploration of Dirac electrons in three dimensions with wide chemical tunability.}
\newpage


\begin{center}
\Large{Main Text}
\end{center}

\textit{Introduction} -- Graphene, the canonical two-dimensional (2D) Dirac semimetal, has provided a rich platform to investigate quantum transport phenomena~\cite{QHE_graphene, FQHE_graphene, unconvensionalSC_graphene, graphene_review, Leslie_graphene} as well as a stimulus for technological innovation~\cite{graphene_photodetector, Thermal_graphene}. 
Graphene forms \textit{sp}$^2$ hybridized orbitals by covalent bonding, and the remaining \textit{p}$_z$ orbital realizes a linear energy dispersion with Dirac points protected by undistorted hexagonal geometry. We call this covalent bonding driven Dirac semimetal (DSM) a \textit{covalent-type} DSM. 

Over the past decade, there have been various theoretical proposals and experimental progress in realizing a three-dimensional (3D) analogue of graphene~\cite{Armitage_review, Murakami_2007, Pyroclore_theory, Dirac_propose_2012, Bulk_Dirac_2014} using \textit{spin-orbit type} DSMs.
This type of DSM requires three conditions; ionic bonding character due to the combination of an electronegative metal and an electropositive ligand to introduce band inversion and form nodal-line-shaped band crossing~\cite{band_inversion_review,Leslie_graphene}, strong spin-orbit coupling (SOC) to open a gap for nearly all of the nodal line, and high crystal symmetry to protect Dirac crossings only along high-symmetry lines or at high-symmetry points in the BZ~\cite{Cd3As2_discovery, Na3Bi_discovery, ZrTe5_discovery}.

These requirements restricted the search for spin-orbit type 3D DSMs to materials containing atoms with relatively large electronegativity differences and strong SOC. Furthermore, due to the difference in electronegativity, Dirac points are formed by bands of different orbital character, leading to asymmetric band crossings. For example, the Dirac crossing in \NaBi{} (in Cd$_3$As$_2$) is created by bands of Na-3\textit{s}/Bi-6\textit{p} (Cd-5\textit{s}/As-4\textit{p}) character and $E_\mathrm{D}$ is limited to $100\,$meV~\cite{Cd3As2_band, Na3Bi_band_calculation}. All these characteristics are unlike 2D graphene, which has very weak spin-orbit interactions~\cite{graphene_weakSOC}, a Dirac dispersion with large $E_\mathrm{D}$, and mostly symmetric conduction / valence bands~\cite{graphene_review, Leslie_graphene}.

\begin{figure}[htb]
  \begin{center}
		\includegraphics[clip, trim=0cm 0cm 0cm 0cm, width=0.6\linewidth]{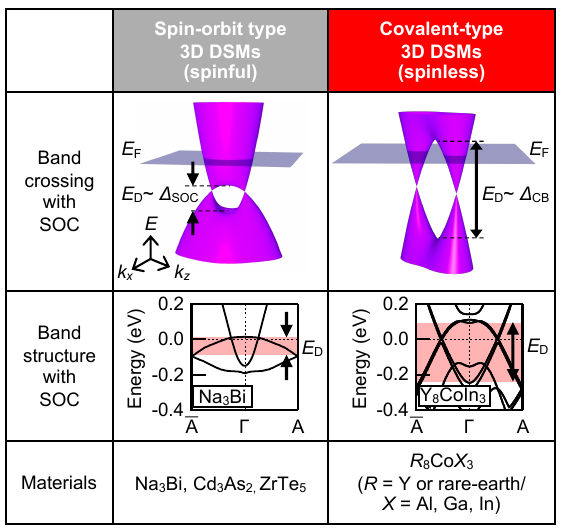}
    \caption[]{(color online). 
Comparison between spin-orbit type and covalent-type three-dimensional Dirac semimetals (DSMs). \textit{top}: In spin-orbit type DSMs, the energy range in which the dispersion is approximately linear (\textit{E}$_{\rm{D}}$) is dominated by the spin-orbit coupling strength ($\Delta_{\rm{SOC}}$), and the conduction and valence bands typically form asymmetric crossings with relatively small and heavy mass, respectively. In contrast, covalent-type DSMs form a clean symmetric band crossing on a large energy scale of the covalent bonding ($\Delta_{\rm{CB}}$). \textit{Middle}: According to $ab$-initio calculations, $E_{\rm{D}}$ of the typical spin-orbit type DSM \NaBi~\cite{Na3Bi_band_calculation} is smaller than that of a recently-proposed covalent-type DSM \YIn~\cite{YIn_theory_2014, Sato_2025}. \textit{Bottom}: Large flexibility in chemical substitution is a key advantage of \RX{}, as compared to other spin-orbit type DSMs. 
}
    \label{Fig1}
  \end{center}
\end{figure}

Covalent-type 3D DSMs, a true analogue of graphene in 3D, were theoretically proposed in 2024~\cite{Er_In_ternary, Ga_series_synthesis,In_series_synthesis, Sato_2025} but remain virtually unexplored experimentally. They form a wide gap between valence and conduction bands due to covalent bonding, with Dirac points protected by crystal symmetry. Covalent-type 3D DSMs fall into a rather unexplored material space, as they combine atoms with similar electronegativity and weak SOC. Moreover, covalent bonding, which has a larger energy scale than SOC, pushes irrelevant bands away from the Fermi energy $E_\mathrm{F}$ due to band hybridization and realizes an extended linear energy dispersion over a wide range of energy (Fig.~\ref{Fig1} (top)). 

\begin{figure}[H]
  \begin{center}
    \includegraphics[clip, trim=0cm 0cm 0cm 0cm, width=0.6\linewidth]{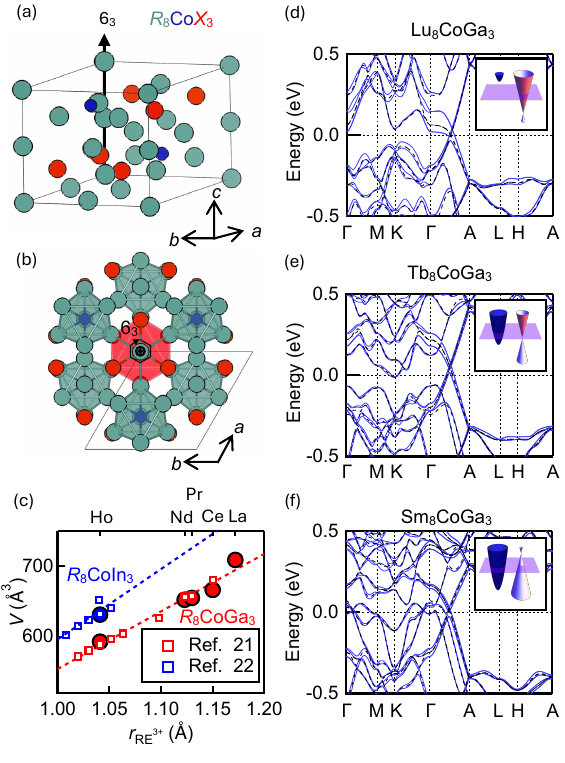}
    \caption[]{
(color online). Crystal and electronic structures of \RX{}. 
(a) Hexagonal crystal structure of the covalent-type Dirac semimetals \RX{} (\textit{R} = Y or rare-earth element; \textit{X} = Ga or In) with the polar space group of $P6_3 mc$. Green, blue, and red spheres represent \textit{R}, Co, and \textit{X} atomic sites, respectively.
(b) Top view of the crystal structure composed of red and green polyhedra with \textit{R}/\textit{X} and \textit{R}/Co atoms, respectively (See Supplementary Fig.~\Ref{Fig_SI_Crystal_structure} for details). The 6$_3$ screw-axis is located at the vertices of the unit cell.
(c) Unit-cell volume as a function of the rare-earth ionic radius, estimated from Rietveld refinement of powder X-ray diffraction data (see Supplementary Fig.~\ref{Fig_SI_Rietveld}~\cite{SI_preprint}). Blue (red) points represent data for \RIn{} (\RGa{}), while open squares are obtained from previous work~\cite{Ga_series_synthesis,In_series_synthesis}. The broken line is a guide to the eye. 
(d–f) Electronic band structures of Lu$_8$CoGa$_3$, Tb$_8$CoGa$_3$, and Sm$_8$CoGa$_3$, respectively, from \textit{ab-initio} calculations with the generalized gradient approximation (GGA). The black-broken and blue-solid lines denote band structures calculated without and with spin-orbit coupling (SOC), respectively. \textit{Inset}: Bandwidth control in \RX{} (schematic). As the ionic radius of $R^{3+}$ increases, the enhanced bandwidth pushes a trivial hole pocket around the \textit{K} point downward, resulting in an upward shift of the Dirac crossing point.
}
    \label{Fig2}
  \end{center}
\end{figure}

To realize covalent-type DSMs in 3D, two main conditions are required: Atoms with similar electronegativity for strong covalent bonding and appropriate crystal symmetry to protect Dirac points from gap opening. \RX{} (\textit{R} = Y or rare earth elements, \textit{X} = Al, Ga, In) has been proposed as a candidate for covalent-type 3D DSMs~\cite{Sato_2025} with similar electronegativity between \textit{R}, Co, and \textit{X} atoms. 
While this \RX{} family had been reported as a topological material candidate in terms of symmetry~\cite{topological_catalogue_2019}, it has not attracted much attention for its covalent bonding character nor for the simplicity of its band structure.
As shown in Figs.~\ref{Fig2}(a) and (b), \RX{} crystallizes in a hexagonal structure with space group $P6_3mc$; the $6_3$ glide crystallographic symmetry along the \textit{c}-axis protects the Dirac crossing point along the $\Gamma \mathchar`-A$ line, i.e., along the \textit{k}$_z$ axis~\cite{Sato_2025}. Thus, the \RX{} satisfy the requirements for covalent-type 3D DSMs and may ultimately overcome some of the limitations of spin-orbit type DSMs.

As shown in Fig.~\ref{Fig1} (middle), \YIn{}, a typical example of the \RX{} family, exhibits a larger $E_{\rm{D}}$ than the representative spin-orbit type DSM \NaBi{}~\cite{Na3Bi_discovery, Na3Bi_band_calculation}. In addition to the difference in energy scale, an electron-hole symmetric gap-opening due to covalent bonding enhances $E_{\rm{D}}$ in \YIn{}, just like in graphene. This contrasts with the asymmetric dispersion in \NaBi{} due to the heavy mass of Bi 6$p$ orbitals.
Importantly -- and in stark contrast to the previously reported spin-orbit type DSMs~\cite{Cd3As2_dope, Cd3As2_dope_Sbdope} -- \RX{} are compatible with a variety of chemical substitutions on \textit{R} and \textit{X} sites as shown in  Fig.~\ref{Fig1} (bottom), offering flexibility in their band structure and magnetic properties. However, the magnetotransport properties of these materials remain unexplored so far. 

In this Letter, we use electronic transport measurements on \RX{} to report the material realization of a new category of 3D DSMs, covalent-type DSMs in the regime of small SOC. We demonstrate $E_\mathrm{F}$ tuning to within only $10 \,$meV from the Dirac point via bandwidth control by chemical substitution.
The carrier mobility increases in inverse proportion to $E_\mathrm{F}$, reaching high values of $3,000\,\mathrm{cm^3/Vs}$ even in polycrystalline samples and evidencing electronic transport governed by Dirac electrons. 
Specifically, when we replace the \textit{R}-site with light rare earth elements, the ionic radius $r_{R^{3+}}$ increases by about $10 \, \%$, as shown in Fig.~\ref{Fig2}(c). The concomitant reduction of the hexagonal lattice constants promotes electron hopping, inducing a larger overlap of conduction and valence bands (see Figs.~\ref{Fig2}(d-f)). We utilize this bandwidth control to tune the $E_\mathrm{F}$ to the Dirac point as depicted in the insets to Figs.~\ref{Fig2}(d-f). 

\begin{figure}[htb]
  \begin{center}
		\includegraphics[clip, trim=0cm 0cm 0cm 0cm, width=0.95\linewidth]{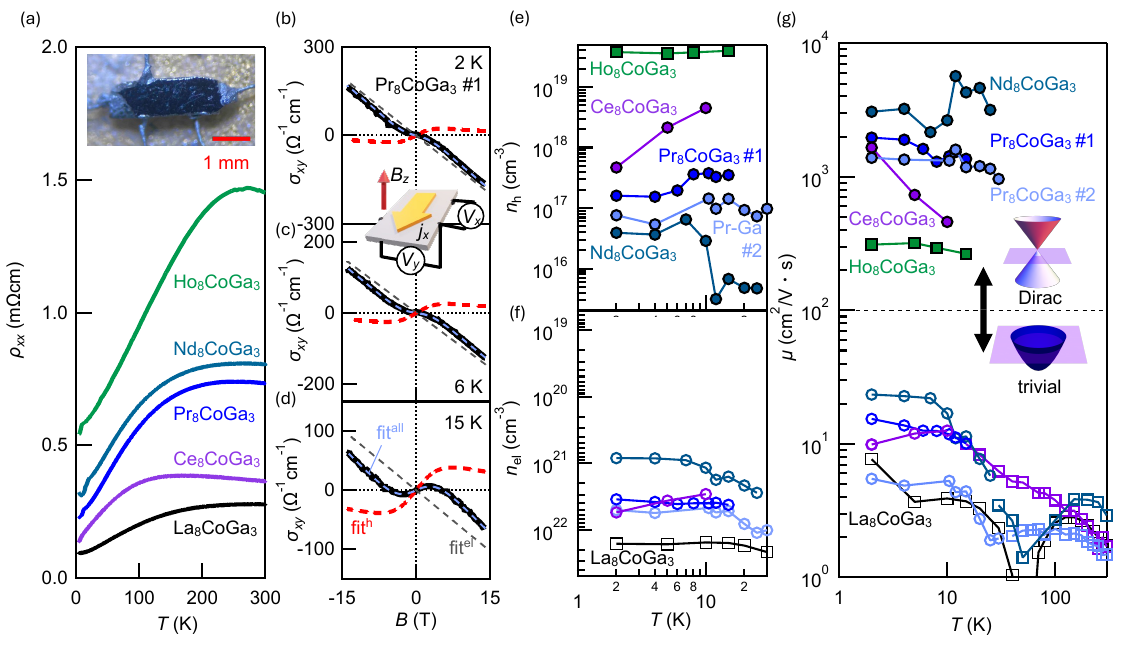}
    \caption[]{
(color online). High-mobility Dirac electrons in electronic transport of \RGa{}. 
(a) Metallic temperature dependence of the longitudinal resistivity $\rho_{xx}$ of \RGa{} (\textit{R}=La, Ce, Pr, Nd, and Ho). \textit{Inset}: optical image of the Ho$_8$CoGa$_3$ sample used.
(b–d) Hall conductivity of Pr$_8$CoGa$_3$ (batch 1) at $2 \,$K, $6 \,$K, and $15 \,$K, respectively. 
Red, blue, and gray broken lines indicate the contributions from high-mobility hole carriers, low-mobility electron carriers, and their sum, respectively, as obtained from a two-carrier model analysis.
(e, f) Carrier density of \RX{} extracted for electron-type and hole-type carriers, respectively. 
(g) Carrier mobility plotted against temperature. We attribute hole-type carriers with high mobility $\mu>100\,$cm$^2$/Vs (filled symbols) to a hole pocket close to a Dirac crossing point and electrons with low mobility $\mu<100\,$cm$^2$/Vs (open symbols) to a trivial pocket around the \textit{K}-point. In panels (e-g), circles (squares) represent the estimation by a two-carrier (single-carrier) model. 
    }
    \label{Fig3}
  \end{center}
\end{figure}

\textit{Experiment} -- \RX{} crystals were obtained by arc-melting and annealing methods~\cite{Ga_series_synthesis,In_series_synthesis}. The crystal structures of the synthesized samples were confirmed by Rietveld refinement of powder X-ray diffraction (XRD) data~\cite{SI_preprint}. The lattice volume evolves in good agreement with previous work (see Fig.~\ref{Fig2}(c))~\cite{Ga_series_synthesis,In_series_synthesis}. Resistivity measurements were carried out using a four-probe method (see inset to Fig.~\ref{Fig3}(a)).

\textit{Results} -- The resistivity of \RGa{} exhibits an overall metallic temperature dependence with a residual resistivity ratio of $2.5\sim3.2$ (see Fig.~\ref{Fig3}(a)). 
The absolute value of the resistivity systematically decreases with rare-earth ionic radius, in accord with the expected bandwidth expansion.  

To extract transport parameters such as the carrier density and carrier mobility, we measure the longitudinal ($\rho_{xx}$) and Hall ($\rho_{yx}$) resistivity, and obtain the Hall conductivity as $\sigma_{xy} = \rho_{yx} / (\rho_{xx}^2 + \rho_{yx}^2)$. In Pr$_8$CoGa$_3$, $\sigma_{xy}$ shows a small positive peak at low fields ($B > 0$) and takes negative values at higher magnetic fields. This sign change in $\sigma_{xy}$ indicates parallel conduction of hole-type and electron-type carriers~\cite{Ziman}. We note that the characteristic behavior of $\sigma_{xy}$ persists even above the magnetic transition temperature ($T_N \approx 12 \,\rm{K}$), indicating a minor contribution from magnetism.



In the semiclassical two-carrier model, $\sigma_{xy}(B)$ is expressed as
\begin{align}
\sigma_{xy}(B) = \frac{n_{\mathrm{h}} e \mu_\mathrm{h}^2 B}{1 + (\mu_\mathrm{h} B)^2} - n_\mathrm{el} e \mu_\mathrm{el}^2 B,
\end{align}
where $n_\mathrm{h}$ and $\mu_\mathrm{h}$ ($n_\mathrm{el}$ and $\mu_\mathrm{el}$) denote the carrier density and mobility of the hole-type (electron-type) carriers, respectively. We assume the mobility of electron-type carriers is sufficiently low ($\mu_\mathrm{el} B \ll 1$), and estimate $n_\mathrm{el}$ and $\mu_\mathrm{el}$ by also considering the magnitude of $\rho_{xx}$~\cite{SI_preprint}. We perform similar analysis for other \RGa{}, as detailed in Fig.~\ref{Fig_SI_Gxy}~\cite{SI_preprint}.

Figures~\ref{Fig3}(e) and (f) show the extracted hole and electron carrier densities $n_\mathrm{h}$ and $n_\mathrm{el}$. In particular, $n_\mathrm{h}$ of (Ce,Pr,Nd)$_8$CoGa$_3$ is on the order of $10^{16}$-$10^{17}\,$cm$^2/$Vs, which is smaller than $n_\mathrm{el}$ by about four orders of magnitude. 
The carrier mobility also shows a large difference between hole-type and electron-type carriers: $\mu_\mathrm{h}$ is about $300 \,$cm$^2/$Vs for Ho$_8$CoGa$_3$, but it exceeds $1,000 \,$cm$^2/$Vs for (Ce,Pr,Nd)$_8$CoGa$_3$ even in polycrystalline samples. We attribute the hole-type carriers with high mobility to a small Fermi surface close to the Dirac points. 

\begin{figure}[htb]
  \begin{center}
		\includegraphics[clip, trim=0cm 0cm 0cm 0cm, width=0.6\linewidth]{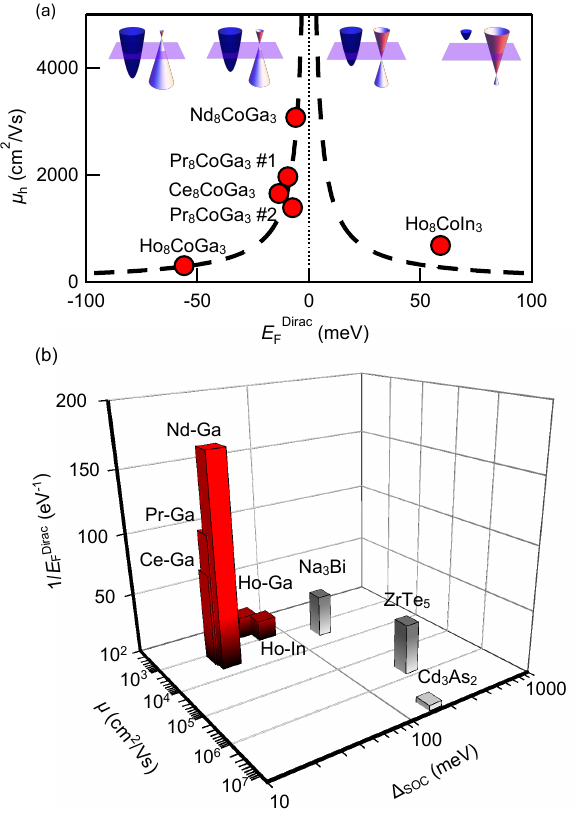}
    \caption[]{
(color online). Covalent-type, three-dimensional Dirac semimetals \RX{} in comparison with other spin-orbit type DSMs.
(a) Carrier mobility plotted as a function of the Fermi energy from the Dirac point, $E_\mathrm{F}{}^{\mathrm{Dirac}}$. The black dotted line is obtained by a fit considering the mobility of a linear Dirac band (see Supplementary Fig.~\ref{Fig_SI_mu_EF}~\cite{SI_preprint}).
\textit{Inset}: Bandwidth expansion induced by the increase in ionic radius (schematic). 
(b) Novel covalent-type Dirac semimetals \RX{} in the regime of small spin-orbit coupling. \RX{} has Dirac electrons tuned close to the Fermi energy, as compared to other spin-orbit type Dirac semimetals.
}
    \label{Fig4}
  \end{center}
\end{figure}

We summarize the filling dependence of $\mu_\mathrm{h}$ of \RGa{}, as well as Ho$_8$CoIn$_3$ in Fig.~\ref{Fig4}(a). Here, we convert $n_\mathrm{H}$ to the Fermi energy measured from the Dirac point, $E_\mathrm{F}{}^{\mathrm{Dirac}}$, assuming an isotropic Fermi surface and using the Fermi velocity $v_{\mathrm{F}}{}^{\mathrm{Dirac}}$ from the \textit{ab-initio} calculations~\cite{SI_preprint}. The mobility $\mu_\mathrm{h}$ of various \RX{} shows an increasing trend toward $E_\mathrm{F}{}^{\mathrm{Dirac}} = 0$.

\textit{Discussion} -- We first examine the dependence of $E_\mathrm{F}{}^{\mathrm{Dirac}}$ on the ionic radius $r_{R^{3+}}$. Except for Ho$_8$CoGa$_3$, the estimated $E_\mathrm{F}{}^{\mathrm{Dirac}}$ monotonically decreases with increasing ionic radius in the order of Ho$_8$CoIn$_3$ and (Nd/Pr/Ce)$_8$CoGa$_3$, as shown in Fig.~\ref{Fig4}(a), Ho$_8$CoGa$_3$ is likely to have a different magnitude of deficiency. The systematic $r_{R^{3+}}$ dependence of $E_\mathrm{F}{}^{\mathrm{Dirac}}$ indicates the successful tuning of the Fermi energy to the Dirac crossing point via bandwidth control in \RX{} (see the insets to Fig.~\ref{Fig4}(a)).

Next, we elaborate on the relationship between $\mu_\mathrm{h}$ and $E_\mathrm{F}{}^{\mathrm{Dirac}}$.  In our experiments, these quantities are found to be inversely proportional to each other, which is well understood in the context of the linear energy dispersion near the Dirac crossing points: The mobility of charge carriers with a linear band dispersion is given by
\begin{equation}
\mu = \frac{e \tau}{m^*}=\frac{e \tau  (v_{\mathrm{F}}{}^{\mathrm{Dirac}})^2}{ E_{\mathrm{F}}{}^{\mathrm{Dirac}}},
\label{eq:mu_linear}
\end{equation}
since the effective mass is expressed as $m^* = \hbar k_{\mathrm{F}}{}^{\mathrm{Dirac}} / {v_{\mathrm{F}}{}^{\mathrm{Dirac}}}$, considering momentum conservation. Here, $\hbar$, $k_{\mathrm{F}}$ and $\tau$ are Planck's constant divided by $2\pi$, the Fermi wave number, and the carrier relaxation time, respectively. The inverse proportionality between $\mu_\mathrm{h}$ and $E_\mathrm{F}{}^{\mathrm{Dirac}}$ is consistently explained by Eq.~(\ref{eq:mu_linear}), supporting the existence of Dirac fermions with a linear band dispersion in \RX{}.

 \vspace{3mm}
\textit{Conclusions} -- {In summary, we demonstrate bandwidth control by \textit{R}-site substitution for the covalent-type, three-dimensional Dirac semimetals \RX{}. We observe parallel conduction of Dirac-like and trivial electrons by electronic transport measurements, and successfully extract the transport properties of the Dirac electrons by fits to a two-carrier model. The Fermi energy is successfully tuned by increasing the ionic radius $r_{\mathrm{R}^{3+}}$ of \textit{R} atoms; the carrier mobility is enhanced as the inverse of the Fermi energy measured from the Dirac point, consistent with the linear-in-energy character of the band dispersion. In Fig.~\ref{Fig4}(b), we compare the observed physical parameters of our \RX{} compounds, which are covalent-type, three-dimensional DSMs, with representative spin-orbit type DSMs~\cite{Na3Bi_band_calculation, Cd3As2_band, ZrTe5_transport, ZrTe5_band}. Our experiments establish the existence of a new class of DSMs in a regime of weak spin–orbit coupling that has so far received little attention, opening up a promising direction beyond the well-studied spin–orbit type topological metals~\cite{Topological_semimetals_prediction_2016,topological_materials_discovery_2019}. 
Combined with the covalent-type nodal-line topological semimetals in square-net with more dense 4$^4$ net materials which possess half-filled \textit{p}$_x$ and \textit{p}$_y$ bands~\cite{topological_in_square_net, delocalized_role_in_square_net, square_net_topology_2022, SrMnBi2_Dirac_fermions, ZrSiS_Dirac_fermions}, covalent-type topological semimetals can be ideal platforms for systematic investigation of topological phenomena with wide chemical and electronic tunability.}  
\vskip\baselineskip
\textbf{Acknowledgements}\\
We acknowledge A. Kikkawa, Y. Onuki, A. Kitaori, and Leslie M. Schoop for fruitful discussions. In this research, we used the ARIM-mdx data system~\cite{hanai2024arim}. This work was supported by JSPS KAKENHI Grant Nos. JP22H04463, JP23H05431, JP22F22742, JP22K20348, JP23K13057, JP24H01607, JP25K17336, and JP24H01604, as well as from the Yazaki Memorial Foundation for Science and Technology and ENEOS TonenGeneral Research/Academic Foundation. This work was partially supported by the Japan Science and Technology Agency via JST CREST Grant Number JPMJCR20T1 (Japan) and JST FOREST (JPMJFR2238). It was also supported by Japan Science and Technology Agency (JST) as part of Adopting Sustainable Partnerships for Innovative Research Ecosystem (ASPIRE), Grant Number JPMJAP2426.\\

\newpage
\bibliography{R8CoGa3}
\newpage

\clearpage
\renewcommand{\thefigure}{S\arabic{figure}}
\setcounter{figure}{0} 

\renewcommand\theequation{S\arabic{equation}}
\setcounter{equation}{0} 

\renewcommand{\thetable}{S\arabic{table}}

\renewcommand\thesection{S \Roman{section}} 
\setcounter{section}{0}

\begin{widetext}
\section*{Supplementary Information}

\section{Crystal growth}
\label{Note_growth}
Precursors of \RX{} are prepared by arc melting stoichiometric ratios of \textit{R} ($99.9 \, \%$, Nippon Yttrium), Co ($99.9 \, \%$, Fujifilm), and \textit{X} (Ga: 99.9999$ \, \%$, Nirako, In:99.9999$ \, \%$, Osaka Asahi Metal Mfg.) in an argon atmosphere after careful evacuation to $10^{-3}\,$Pa. The ingots are turned and remelted at least three times to ensure homogeneity, and the weight loss is estimated to be $0.5 \,\%$. The obtained precursors are cracked into small pieces with a size of $0.5 \sim 2 \,$mm, which are wrapped in Ta foil. The wrapped sample is sealed in an evacuated silica tube and heated in an electric furnace. For (La/Ce/Nd/Pr)$_8$CoGa$_3$ compounds, samples are kept at 400$\,{}^\circ$C for two weeks~[S1], while Ho$_8$Co(Ga/In)$_3$ samples are kept at 600$\,{}^\circ$C for two months~[S2].

\section{Rietveld analysis}
We measure powder X-ray diffraction (XRD) of crushed \RX{} samples at room temperature using a commercial Rigaku SmartLab diffractometer with Cu-$K_\alpha$ radiation (wavelength $\lambda = 1.5406 \, \AA$). The crystal structure refinement is performed using the FullProf software suite~[S3]. 
The powder pattern is refined using the hexagonal crystal structure $P6_3mc$ (Space group No. 186), as shown in Supplementary Fig.~\ref{Fig_SI_Rietveld}. The extracted lattice volumes are plotted in Fig.~\ref{Fig2}(c) and refined structural parameters are listed in Table~\ref{SI_table}.

\section{Resistivity measurement}

We perform resistivity measurements on polycrystalline $R_8$Co(Ga/In)$_3$ using samples polished into a rectangular shape with dimensions $\sim 1 \,$mm, in a Quantum Design Physical Properties Measurement System (PPMS). We measure both longitudinal $\rho_{xx}$ and Hall resistivity $\rho_{yx}$ by a standard four-probe method. We symmetrized (antisymmetrized) $\rho_{xx}$ ($\rho_{yx}$) with respect to the magnetic field. We obtain the Hall conductivity as $\sigma_{xy} = \rho_{yx} / (\rho_{xx}^2 + \rho_{yx}^2)$. 

Supplementary Figures~\ref{Fig_SI_Gxy}(a-f) show $\sigma_{xy}$ of (La/Ce/Pr/Nd/Ho)$_8$CoGa$_3$ and Ho$_8$CoIn$_3$, respectively. We successfully performed a two-carrier fit for (Ce/Pr/Nd)$_8$CoGa$_3$, however, the monotonic, nearly $B$-linear, field dependence of $\sigma_{xy}$ of La$_8$CoGa$_3$, likely originating from the low carrier mobility of Dirac electrons in this sample with large $\left|E_\mathrm{F}\right|$, prohibits a proper analysis with a two-carrier model. Thus, we use a single-carrier model to estimate the carrier density and mobility of La$_8$CoGa$_3$. As for Ho$_8$CoGa$_3$ and Ho$_8$CoIn$_3$, the Hall conductivity $\sigma_{xy}$ is well reproduced by the single carrier model without adding contribution from another carrier, as shown in Supplementary Fig.~\ref{Fig_SI_Gxy}(f). Therefore, we used a single carrier model for the estimation of carrier density and mobility for these two compounds. 

\newpage
\begin{figure}[htb]
  \begin{center}
    \includegraphics[clip, trim=0cm 0cm 0cm 0cm, width=0.95\linewidth]{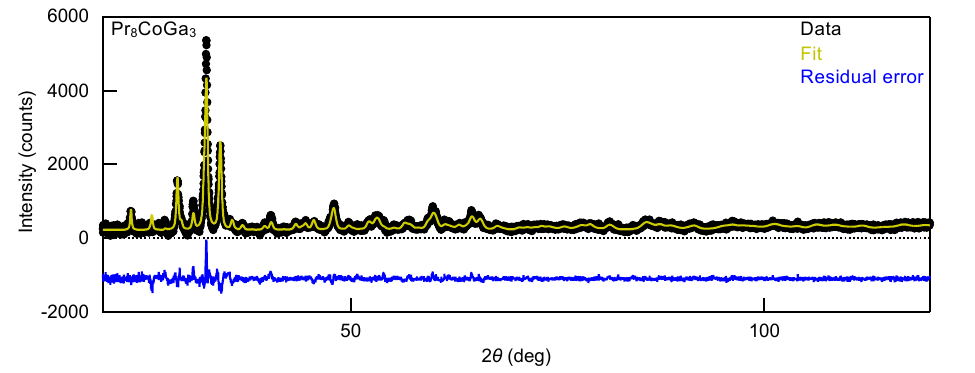}
    \caption[]{
    (color online). Rietveld analysis on the powder X-ray diffraction pattern of Pr$_8$CoGa$_3$ measured at room temperature. Refined parameters are described in the table~\ref{SI_table}.
    }
    \label{Fig_SI_Rietveld}
  \end{center}
\end{figure}
\renewcommand{\arraystretch}{2}

\newpage
\begin{table}[h]
	\centering
	\caption{\textbf{Structural parameters obtained from Rietveld analysis of powder X-ray
diffraction for crushed polycrystalline Pr$\bm{_8}$CoGa$\bm{_3}$.} We confirm the hexagonal space group $P6_3mc$ with lattice constants \textit{a} = 10.4614(3)$\,$\AA{} and \textit{c} = 6.9033(4)$\,$\AA.} 
	\label{SI_table} 
	\begin{tabular}{|c|r|r|r|r|r|r|r|} \hline
		Atom&Coordinates&x&y&z&Occupancy&Site&Symmetry\\ \hline \hline
		Pr1&(x, -x, 0)&0.1856(4)&-0.1856(4)&0&0.50000&6c&.m.\\\hline
		Pr2&(1/3, 2/3, z)&1/3&2/3&0.619(6)&0.16667&2b&3m.\\\hline
		Pr3&(x, -x, z)&0.4619(3)&-0.4619(3)&0.284(2)&0.50000&6c&.m.\\\hline
		Pr4&(0, 0, z)&0&0&0.280(5)&0.16667&2a&3m.\\\hline
            Ga1&(x, -x, z)&0.1611(6)&-0.1611(6)&0.545(2)&0.50000&6c&.m.\\\hline
            Co1&(1/3, 2/3, z)&1/3&2/3&0.04(4)&0.16667&2b&3m.\\\hline
\end{tabular}
\end{table}

\newpage
\begin{figure}[H]
  \begin{center}
    \includegraphics[clip, trim=0cm 0cm 0cm 0cm, width=0.95\linewidth]{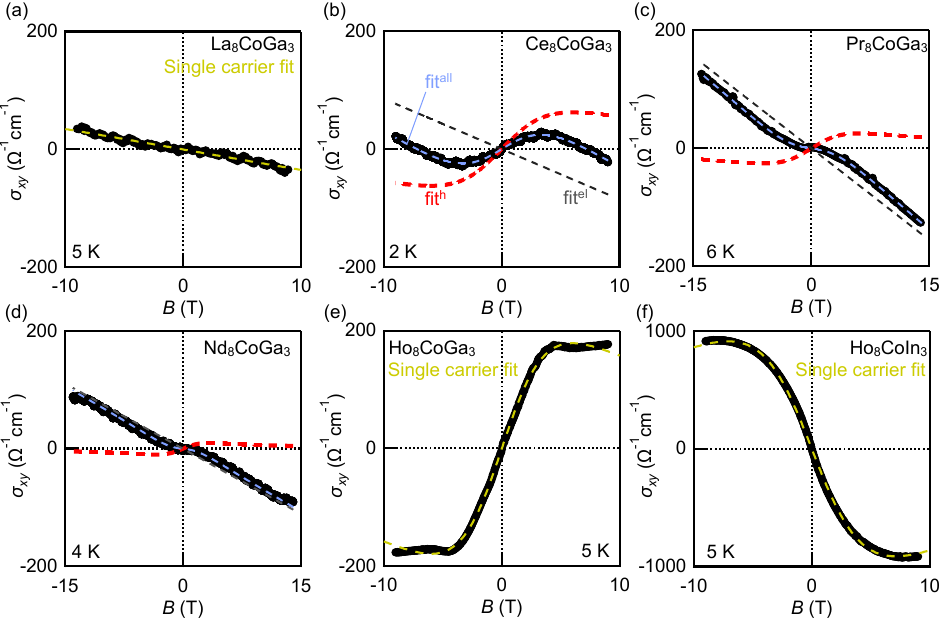}
    \caption[]{
    (color online). Hall conductivity $\sigma_{xy}$ of (La/Ce/Pr/Nd/Ho)$_8$CoGa$_3$ and Ho$_8$CoIn$_3$.}
    \label{Fig_SI_Gxy}
  \end{center}
\end{figure}

\newpage
\begin{figure}[H]
  \begin{center}
    \includegraphics[clip, trim=0cm 0cm 0cm 0cm, width=0.95\linewidth]{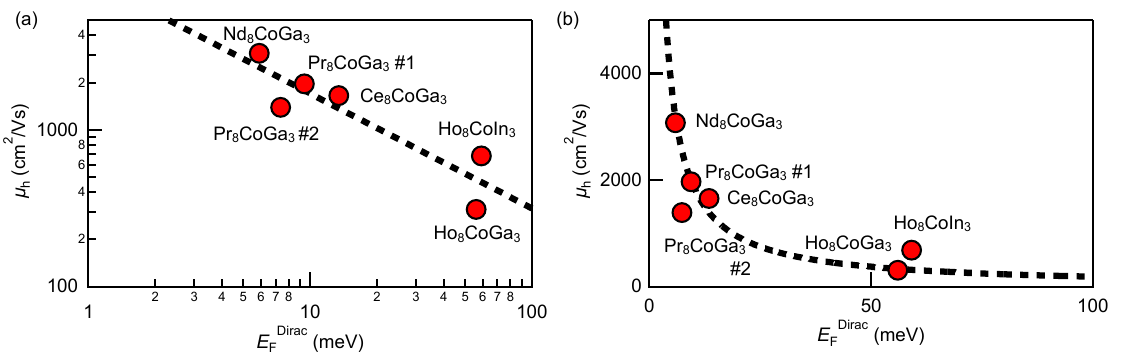}
    \caption[]{
    (color online). (a) Logarithmic plot of carrier mobility and of the absolute value of the Fermi energy from the Dirac point $E_{\rm{F}}{}^{\rm{Dirac}}$. Black dashed line: linear fit to the data. 
    (b) Carrier mobility against $E_{\rm{F}}{}^{\rm{Dirac}}$. Black dashed line: fit to the expected $\sim 1/ E_\mathrm{F}$ behavior. From the fitting based on Eq.~(\ref{eq:mu_linear}), we estimate the relaxation time as $\tau\sim9.0\times10^{-14}\,$s. This value is slightly smaller than the typical value of single-crystalline topological semimetals.
    }
    \label{Fig_SI_mu_EF}
  \end{center}
\end{figure}

\newpage
\begin{figure}[H]
  \begin{center}
    \includegraphics[clip, trim=0cm 0cm 0cm 0cm, width=0.95\linewidth]{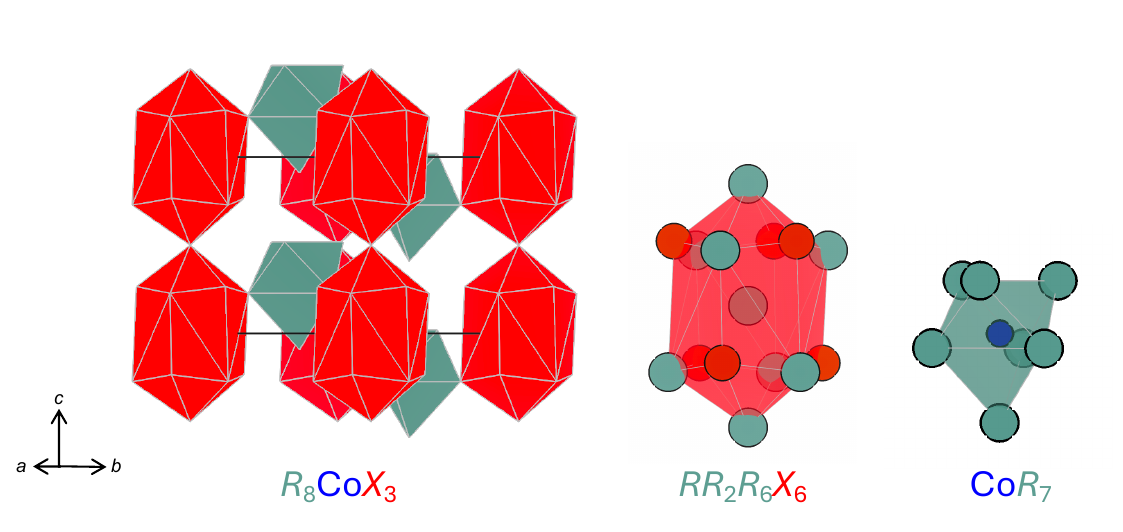}
    \caption[]{
    (color online). Hexagonal \RX{} crystal structure from the view angle perpendicular to the \textit{c}-axis. This structure can be seen as a successive substitution of the atoms in the CaIn$_2$ structure ($P6_{3}/mmc$) with various polyhedral units~[S2]. Ca and In sites of CaIn$_2$ are replaced with $R$-centered hexagonal prism with two additional $R$-atoms on top and bottom of the prism [$RR_2R_6X_6$] and distorted octahedra containing Co-atom with an additional $R$-atom [Co$R_{7}$], respectively. As a result, the relation between the structure of CaIn$_2$ and \RX{} can be expressed as follows: $\mathrm{CaIn}_{2} = 2\mathrm{Ca} + 4\mathrm{In} = 2[RR_{2}R_{6}X_{6}]-2R + 4[CoR_{7}] - 12R = 32R + 4\mathrm{Co} + 12X = 2$\RX{}. The subtraction of $2R$ and $12R$ is introduced to take into account the $R$-atoms shared by two fragments.
    }
    \label{Fig_SI_Crystal_structure}
  \end{center}
\end{figure}
\clearpage
\newpage
\noindent[S1] Yu. N. Grin', O.M. Sichevich, R.E. Gladyshevskij, and Ya.P. Yarmolyuk, Kristallografiya \textbf{29}, 708 (1984).

\noindent[S2] M. Dzevenko, I. Bigun, M. Pustovoychenko, L. Havela, and Y. Kalychak, Intermetallics \textbf{38}, 14 (2013).

\noindent[S3] J. Rodr{\'i}guez-Carvajal, Physica B: Condens. Matter \textbf{192}, 55 (2013).
\end{widetext}
\end{document}